\documentclass{icrc2009}

\usepackage{graphicx}   % for including figures
\usepackage[caption=false]{caption}    % for captions
\usepackage[font=footnotesize]{subfig} % subfig.sty for a double column floating figure using two subfigures
\usepackage{fixltx2e}
\usepackage{url}

\newcommand{\shorttitle}[1]%
{\markboth{Proceedings of the 31\MakeLowercase{$^{st}$} ICRC, {\L}\'{o}d\'{z} 2009}{#1} }
\newcommand{\etal}{\MakeLowercase{\textit{et al. }}} % "et al."

\begin{document}
\title{Multiwavelength observations of a TeV-Flare from W Com}

\author{\IEEEauthorblockN{
      Gernot Maier\IEEEauthorrefmark{1} for the VERITAS collaboration\IEEEauthorrefmark{2} \\
      Elena Pian\IEEEauthorrefmark{3} for the AGILE collaboration\IEEEauthorrefmark{4} }
			  
                            \\
\IEEEauthorblockA{\IEEEauthorrefmark{1}Department of Physics, McGill University, H3A 2T8 Montreal, QC, Canada (maierg@physics.mcgill.ca)}
\IEEEauthorblockA{\IEEEauthorrefmark{2} see R.A. Ong et al (these proceedings) or http://veritas.sao.arizona.edu/conferences/authors?icrc2009}
\IEEEauthorblockA{\IEEEauthorrefmark{3} Osservatorio Astronomico di Trieste, Via G.B. Tiepolo 11, I-34143 Trieste, Italy}
\IEEEauthorblockA{\IEEEauthorrefmark{4} see http://agile.rm.iasf.cnr.it/index.php for a full author list}
}

% please write the preseter's name and short title (3-4 words maximum)
%    which will appear at the header of the even pages.
\shorttitle{G.Maier \etal observations of a TeV-Flare from W Com}
\maketitle

\begin{abstract}
We report results from an intensive campaign of multiwavelength observations of the intermediate-frequency-peaked 
BL Lacertae object W Com (z=0.102) during a strong outburst of very high energy (VHE; E $>$ 180 GeV) 
gamma-ray emission in June 2008. 
The initial detection of this VHE flare by VERITAS, an array of four 12-m diameter imaging atmospheric-Cherenkov telescopes, 
was followed by observations in high-energy gamma rays (AGILE, E $>$ 100 MeV), X-rays (\emph{Swift}), optical and radio wavelengths.
 The VHE gamma-ray signal was detected by VERITAS on June 7-8 with a flux about three times brighter than 
 during the discovery of VHE gamma-ray emission from W Com by VERITAS in March 2008. 
 A detailed study of the spectral energy distribution of W Com during this flare, including theoretical work, will be presented.
 \end{abstract}

\begin{IEEEkeywords}
gamma-ray observatory; blazars; W Com
\end{IEEEkeywords}
 
\section{Introduction}
 
W Com is an intermediate-frequency-peaked BL Lac (IBL) at a redshift of $z=0.102$. 
The object was discovered in high-energy gamma rays (from 100 MeV to 10 GeV) by EGRET \cite{Hartman1999}
and in
very high energy gamma rays in March 2008  during a strong outburst
of about 4 days duration by VERITAS \cite{Acciari-2008}. 
The broadband spectral energy distribution (SED) of W Com is characterized by two peaks, arising from 
synchrotron radiation at radio to X-ray frequencies and high energy emission due to 
inverse Compton scattering or hadronic interactions at hard X-ray to gamma-ray energies.
The X-ray and gamma-ray emission from W Com has been found to be variable on a 
timescale of hours to days \cite{Massaro-1999}.

Here we present contemporaneous observations of W Com during a second strong flare in June, 2008 with
VERITAS, AGILE, and \emph{Swift}.
  
\section{Observations and Results}

\subsection{VERITAS}

VERITAS is an array of four imaging atmospheric-Cherenkov telescopes
located at the Fred Lawrence Whipple Observatory
in southern Arizona.
It combines a large effective
area (up to $10^5$ m$^2$) over a wide energy range (100 GeV
to 30 TeV) with good energy (15-20\%) and angular
($\approx0.1^{\mathrm{o}}$) resolution.
The field of view of the VERITAS telescopes is 3.5$^{\mathrm{o}}$.
The high sensitivity of VERITAS enables
the detection of sources with a flux of 1\% of the Crab
Nebula in less than 50 hours of observations.
For more details on the VERITAS
instrument, see e.g.~\cite{Ong-2009}.

\begin{figure}[!t]
\centering
\includegraphics[width=3.0in]{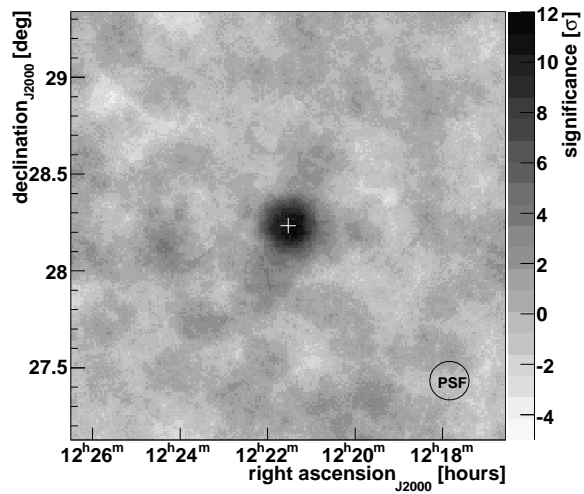}
\caption{
VERITAS significance map of the region around W Com 
for MJD 54624.16 to  54625.24.
The background is estimated using the reflected region model.
The position of W Com derived from radio data is indicated by a  white cross.
The circle at the bottom right indicates the angular resolution of the 
VERITAS observations.
}
\label{fig:VERITASSky}
\end{figure}

VERITAS observed the sky around W Com for about four hours in June 2008.
All observations passed run quality selection 
criteria, which remove data taken during bad weather or with
hardware-related problems.
Data were taken in wobble mode, wherein the source was positioned at a fixed
offset from the camera center.
Parts of the observations were undertaken in moonlight conditions, where 
the elevated background light levels lead to a lower flux sensitivity and higher energy threshold for the
detection of $\gamma$-rays.
The different elevations  combined with
the continuously changing background light conditions due to the moon 
results in a wide range of energy thresholds from 180 to 420
GeV\footnote{
The energy threshold is defined as the peak of the differential counting rate
for a Crab Nebula-like spectrum.} for the present observations.
Table \ref{table:VERITASobservations} lists observation times, elevation range,
and background light conditions.

  \begin{table*}[!h]
  \caption{Details of the VERITAS observations of W Com in 2008 June. Fluxes and upper flux limits are given at the 99\% 
  confidence level 
  for an energy threshold of 200 GeV.}
  \label{table:VERITASobservations}
  \centering
  \begin{tabular}{|c|c|c|c|c|c|}
  \hline
   MJD  &  elevation & observation & average pedestal & significance & flux or upper flux limit \\
             &    range  &     time    [min]               & variations [dc]\footnotemark & [$\sigma$] & [cm$^{-2}$s$^{-1}$]\\
   \hline 
 54624.16 - 54624.23 & 53-73$^{\mathrm{o}}$ & 100.2 & 7.8-8.0 & 8.9 & $(5.0\pm0.8)\times 10^{-11}$ \\
54625.17 - 54625.24 & 49-68$^{\mathrm{o}}$ & 100.2 & 8.1-9.7 & 7.9 & $(6.2\pm1.2)\times 10^{-11}$ \\
54626.18 - 54626.20 & 59-60$^{\mathrm{o}}$ & 32.0  & 12.2-12.3 & -1.0 & $<3.15\times 10^{-11}$ \\  
  \hline
  \end{tabular}
  \end{table*}
  \footnotetext{The average pedestal variations in digital counts [dc] describe the increased background light levels due to the moon.
Values of 6.5 to 6.8 are typical for regular observations of extragalactic targets in moonless nights.}

The VERITAS data analysis steps consist of image calibration and cleaning,
second-moment parameterization of these images \cite{Hillas-1985},
stereoscopic reconstruction of the event impact position and direction,
gamma-hadron separation, spectral energy reconstruction
(see e.g.~\cite{Krawczynski-2006})
as well as the generation of photon sky maps.
The majority of the far more numerous background events are rejected by comparing the shape of the event images in each 
telescope with the expected shapes of gamma-ray showers modeled by Monte Carlo 
simulations\footnote{The varying  brightness levels due to moon are taken into account in the Monte Carlo simulations.}.
These \textit{mean-reduced-scaled width} and \textit{mean-reduced-scaled length}
cuts (see definition in \cite{Krawczynski-2006}), and an additional cut on the arrival direction of the incoming gamma ray ($\Theta^2$)
reject more than 99.9\% of the cosmic-ray background while keeping 45\% of the gamma rays.
The cuts applied here are:
integrated charge per image $>$ 400 digital counts ($\approx$75 photoelectrons), 
$-1.2 < $ mean-reduced-scaled width/length $< 0.5$, and \mbox{$\Theta^2 < 0.015$ deg$^2$}.
The background in the source region is estimated from the same field of view
using the ``reflected-region'' model \cite{Aharonian-2001}.

VERITAS detected a significant flux of high-energy gamma rays from W Com for the entire data set  (June 7-9 2008).
A total of 117 excess events (195 ON events and 78 normalized OFF events, normalization factor $\alpha=0.10$)
have been measured.
This corresponds to a total significance of 10.3 standard deviations, calculated following equation 17 in  \cite{Li-1983}.
The daily significances and fluxes above 200 GeV can be found in Table \ref{table:VERITASobservations}.
W Com was not detected on June 9 2008 (MJD 54626), but observations were restricted to 32 min due to
bright background light levels caused by the moon.
The sky around W Com in high-energy gamma rays for the June 7-8 2008 observations
  is shown in Figure \ref{fig:VERITASSky}.
The region around the TeV-blazar 1ES 1218+304 \cite{Acciari-2009b}, located about $2^{\mathrm{o}}$ north
of W Com, has been excluded from the background estimation.

\begin{figure}[!t]
\centering
\includegraphics[width=3.0in]{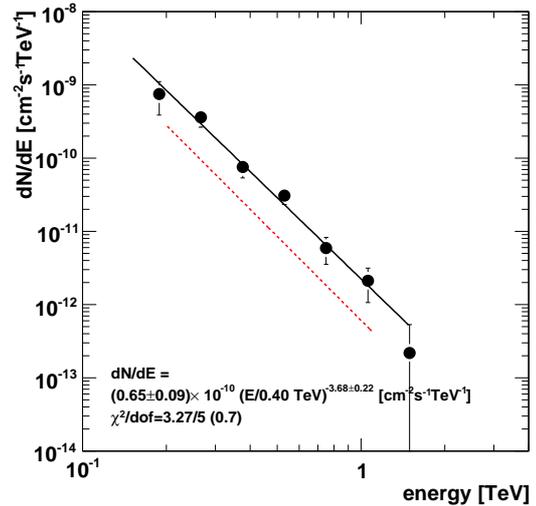}
\caption{
Differential energy spectrum for W Com for MJD 54624.16 to  54625.24 (June 7-8 2008).
The markers indicate measured data points, the continuous line a fit 
assuming a power-law distribution.
The spectral energy distribution measured by VERITAS in March 2008 
\cite{Acciari-2008} is indicated by a dashed line.
}
\label{fig:VERITASEnergySpectrum}
\end{figure}

The differential photon spectra between 180 GeV and 3 TeV
 for the combined measurements from June 7 and 8 2008 are shown in Figure \ref{fig:VERITASEnergySpectrum}.
The shape is consistent with a power law $\mathrm{dN/dE = C\times(E/0.4GeV)}$$^{-\Gamma}$ with a photon index   
$\Gamma = 3.68\pm0.22_{stat}$ and a flux normalization constant C = $(6.5\pm0.9_{stat})\times10^{-11}$ 
cm$^{-2}$s$^{-1}$TeV$^{-1}$.
The $\chi^2$ of the fit is 3.27 for 5 degrees of freedom\footnote{see \cite{Acciari-2009a} for a discussion of 
systematic errors.}.
For comparison, the flare in high-energy gamma rays from W Com in March 2008 is well fit
by the same power law with $\Gamma = 3.81\pm0.35_{stat}$ and
C = $(2.00\pm0.31_{stat})\times10^{-11}$ 
cm$^{-2}$s$^{-1}$TeV$^{-1}$.

 \subsection{AGILE}

The Gamma-Ray Imaging Detector (GRID, 30 MeV - 30 GeV) onboard the high
energy astrophysics satellite AGILE (\cite{Tavani2008a, Tavani2008b}) observed
W Com uninterruptedly from 9 (18:00 UT) to 15 (12:00 UT) June, 2008.
Preliminary and partial results of this observation were reported in \cite{Verrecchia2008}.
The GRID data were analyzed using the AGILE standard pipeline \cite{Vercellone2008}, with a bin
size of $0.25^{\mathrm{o}} \times 0.25^{\mathrm{o}}$. Only events flagged as gamma rays and not recorded while
the satellite crossed the South Atlantic Anomaly were accepted.
We also rejected all events with reconstructed direction within $10^{\mathrm{o}}$ from the Earth limb,
thus reducing contamination from Earth's gamma-ray albedo.
W Com, observed at about 3 degrees off-axis with respect to the boresight, was only detected at
the 3.7-$\sigma$ level ($E>100$ MeV) from 12 (03:00 UT) to 13 (03:00 UT) June, 2008 with a flux
of about $90 \times 10^{-8}$ ph~s$^{-1}$~cm$^{-2}$, similar to the flux detected by EGRET \cite{Hartman1999}.
In the rest of the observing period the source was not detected above $3 \sigma$;
we report the measurement and upper limits (95\% confidence level) in Table \ref{table:AGILEobservations} and Figure \ref{fig:SED}.
The paucity of photons prevents us from extracting a spectrum.  
SuperAGILE, the hard X-ray imager onboard AGILE (20-60 keV, \cite{Feroci2007})
observed the source for a net exposure time of 253 ks.
The source position in the orthogonal SuperAGILE reference system was $\sim$ (3,0) deg,
which means that  almost the full on-axis effective area was exposed \cite{Feroci2007}.
W Com was not detected, and
we estimate a 3-$\sigma$ upper limit in the 20-60 keV energy of 6 mCrab $\simeq 6.9 \times 10^{-11}$ 
erg~cm$^{-2}$~s$^{-1}$.

 \begin{table}[!h]
 \caption{Details and results of the AGILE observations of \mbox{W Com}. }
 \label{table:AGILEobservations}
 \centering
 \begin{tabular}{|c|c|c|}
 \hline
 MJD & significance & flux or \\
          &                        & $2\sigma$ upper flux limits \\
          &                        & (E$>$100 MeV) [ph/cm$^{2}$/s] \\
  \hline
54626.75 - 54629.12 & $<3\sigma$ & $<40\times10^{-8}$  \\
54629.12 - 54630.12 & $3.7\sigma$ & $(90\pm34)\times10^{-8}$ \\
54630.12 - 54632.50 & $<3\sigma$ & $<37\times10^{-8}$ \\
   \hline
  \end{tabular}
  \end{table}

\subsection{Swift}

Observations of W Com  with the \emph{Swift} satellite were taken on June 8 and 9, 2008
(MJD 54625 and MJD 54626), respectively,  and were partly contemporaneous with VERITAS and AGILE observations.
Presented here are the data from the XRT \cite{Burrows-2005} and UVOT \cite{Poole-2008} instruments. 
A standard data reduction was performed with the HEAsoft 6.5 package.
All XRT observations were taken in Photon Counting (PC) mode. 
The count rate is above 0.5 cts/s during some observations and photon pileup occurred. 
For these observations an annular source extraction region was applied to exclude the central 3 pixels.
See \cite{Acciari-2009a} for a detailed
description of the \emph{Swift} data analysis.

A large data set of optical and radio observations has been taken during the considered time period, but 
is not shown here; see \cite{Acciari-2009a}  for more details.
 
\section{Discussion} 

 \begin{figure*}[th]
  \centering
  \includegraphics[width=6in]{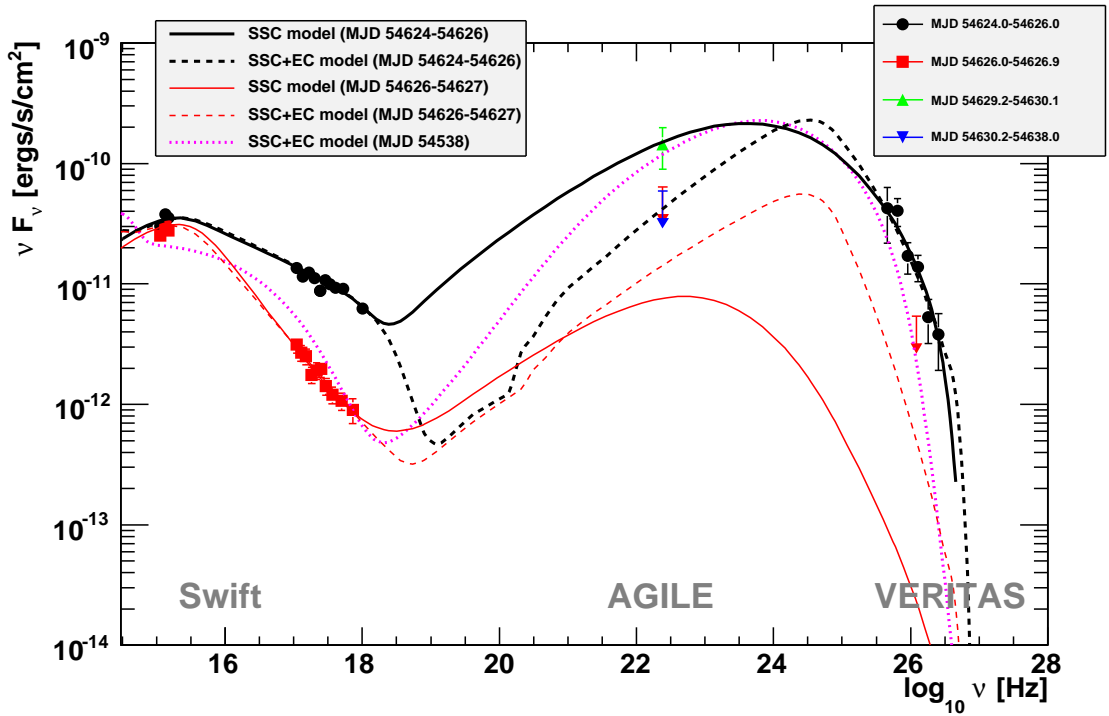}
  \caption{
  Spectral energy distribution of W Com for MJD 54624 to 54638 including
  VERITAS, AGILE, and \emph{Swift} XRT/UVOT data.
  Downwards pointing arrows indicate $2 \sigma$ (AGILE data) or $3 \sigma$ (VERITAS data) upper flux limits \cite{Helene-1983}.
  Details of the SSC and SSC+EC model fits are described in the text.
  The SSC+EC model fit to the March 2008 data (MJD 54538; see \cite{Acciari-2008}) is indicated
  by a dotted line.
    }
  \label{fig:SED}
 \end{figure*}
 
 Figure \ref{fig:SED} shows the spectral energy distribution (SED) of W Com from UV to VHE gamma rays for
 three different time intervals: the large flare measured by VERITAS and \emph{Swift} (June 7-8 2008; MJD 54624 - 54626), 
 AGILE, VERITAS and \emph{Swift} measurements
 with no significant signal above 100 MeV and 200 GeV (June 9 2008; MJD 54626-54627) and data taken by AGILE  after
 MJD 54630 (June 12-15 2008).
 
 We first note from Figures \ref{fig:VERITASEnergySpectrum} and \ref{fig:SED} presented here and Figures 3 and 4 in \cite{Acciari-2008}  
 that while  the integral flux above 200 GeV is about 3 times higher for June 7-8 2008 than for March 2008, no change in the 
 photon index is observed. 
We observe furthermore the approximately  same flux and spectral index in X-rays on June 9 2008 as in March 2008; the short
observation time with VERITAS on June 9 does not allow a detection of VHE gamma rays at a flux level similar to that
observed in March 2008.

The SED of W Com can be fit by a one-zone, time dependent, combined synchrotron self-Compton (SSC) and external Compton (EC) jet model.
The equilibrium version of the code of \cite{Boettcher-2002} has been used.
In this model a population of ultrarelativistic nonthermal electrons and positrons is continuously 
injected into a spherical emitting volume
of comoving radius $R_b$ moving with  a relativistic speed $\beta=(1-1/\Gamma^2)^{-1/2}$ along the jet.
The spectrum of the injected pair population is specified through the injection power $L_e$ and
an unbroken power law with low- and high-energy cutoffs $\gamma_{min}$ and $\gamma_{max}$, respectively, and
a spectral index $q$.
The injected particles suffer energy losses via synchrotron emission and upscattering of synchrotron or external 
photons or may escape from the emitting region.
The particle escape is parameterized through an energy-independent timescale $t_{esc} = \eta R_B / c$, with $\eta \geq 1$.
$\gamma\gamma$-absorption of VHE gamma rays on low-energy extragalactic background photons has been taken
into account using the scenario described in \cite{Franceschini-2008}.
It should be noted that not all parameters of the model are constrained by the available data, and subject to some rather
arbitrary, unconstrained choices (like the Doppler factor and details of the external radiation field).

Some preliminary fit results from the SSC models (solid lines) and SSC+EC (dashed lines) 
are shown in Figure \ref{fig:SED}, for a detailed discussion of the different fit results see \cite{Acciari-2009a} .
The parameters for the SSC model for the June 7-8 2008 observations are: 
$\gamma_{min}=9\times10^3$, $\gamma_{max}=2.5\times10^5$, $q=2.55$, $L_e=3.4\times10^{44}$ erg/s, 
a magnetic field strength of 0.24 G, a Doppler factor $\delta=20$ and a size of the emission region of $R_B=3\times10^{15}$ cm.
The SSC model provides a reasonable fit, but requires a sub-equipartition magnetic field (equipartition parameter
$\epsilon_B=2.3\times10^{-3}$). 
This is similar to what was found in the SSC-modeling of the March 2008 data \cite{Acciari-2008}.
The non-detection of W Com  at energies above 200 GeV on and after June 9 2008
 leaves the model fits  unconstrained and without
unique solution. 
However, all models predict
a significant change in the spectral index of the radiating electrons between high (June 7-8 2008) and
low (June 9 2008) X-ray states, being much softer in the low state ($q=2.55$ vs $q=3.5$).
 
The SSC+EC model includes photons emitted from the disk as a steady-state blackbody radiation peaking in the near-infrared.
The SED is well described by this model with a magnetic field close to equipartition ($\epsilon_B=0.3$).
Figure  \ref{fig:SED}  shows for comparison the SSC+EC fit results to the March 2008 data.
The SSC+EC model fit obtained in March 2008 exceeds the upper flux limit  at 100 MeV
from AGILE observations on  June 7 2008,
while it describes well the optical and X-ray data for this day (see dotted line in Figure \ref{fig:SED}).

The light crossing time $\tau = R_B/(c\times\delta)$ is with values between 90 and 300 min in all models (SSC and SSC+EC) consistent with
variability timescales observed  in X-rays and gamma rays from W Com and other VHE blazars.

\section{Summary}

The intermediate-frequency-peaked BL Lac object W Com was discovered during a strong outburst in March 2008.
Here results from observations on June 7 and 8, 2008 of a second flare, 
with a three times higher flux of gamma rays above 200 GeV, are presented.
The VERITAS observations were followed by radio, optical, UV and X-ray (\emph{Swift}), and gamma ray (AGILE, $E>$100 MeV) observations.
The SED can be modeled by a simple leptonic SSC model, but the wide separation of the SED peaks requires 
 a rather low
ratio of the magnetic field to electron energy density of $\epsilon_B=2.3\times10^{-3}$.
The SSC+EC model returns magnetic field parameters closer to equipartition, providing a satisfactory description of the broadband SED.

The strong variability of W Com at all energies on timescales of days or less shows that only truly contemporaneous data can provide 
serious constraints to the various emission models.
The results presented here show that more data, especially in the hard X-ray to gamma-ray regime, is necessary to understand
particle acceleration and high-energy gamma-ray emission in this interesting object.

\section{Acknowledgments}
This research is supported by grants from the US Department
of Energy, the US National Science Foundation, and the Smithsonian
Institution, by NSERC in Canada, by Science Foundation
Ireland, and by STFC in the UK.
We acknowledge the
excellent work of the technical support staff at the FLWO and
the collaborating institutions in the construction and operation
of the instrument.
EP acknowledges support from the Italian Space Agency through grant ASI I/088/06/0.
We acknowledge the efforts of the Swift team
for providing the UVOT/XRT observations.

\end{document}